\documentclass[lettersize,journal]{IEEEtran}
\usepackage{amsmath,amsfonts}
\usepackage{algorithmic}
\usepackage{array}
\usepackage[caption=false,font=normalsize,labelfont=sf,textfont=sf]{subfig}
\usepackage{textcomp}
\usepackage{stfloats}
\usepackage{url}
\usepackage{verbatim}
\usepackage{graphicx}
\usepackage{color}
\usepackage{ulem}
\def\BibTeX{{\rm B\kern-.05em{\sc i\kern-.025em b}\kern-.08em
    T\kern-.1667em\lower.7ex\hbox{E}\kern-.125emX}}
\usepackage{balance}

\begin{document}

\title{Simultaneous MMC readout using a tailored $\mu$MUX based readout system}

\author{D. Richter, M. Wegner, F. Ahrens, C. Enss, N. Karcher, O. Sander, C. Schuster, M. Weber, T. Wolber, S. Kempf
\thanks{
This work has been performed in the framework of the DFG research unit FOR 2202 (funding under grants En299/7-1, En299/7-2, and En299/8-2).

D. Richter and F. Ahrens are with Kirchhoff Institute for Physics, Heidelberg University.

M. Wegner was with Kirchhoff Institute for Physics, Heidelberg University. He is now with Institute for Data Processing and Electronics and Institute of Micro- and Nanoelectronic Systems, both Karlsruhe Institute of Technology.

C. Schuster was with Kirchhoff Institute for Physics, Heidelberg University. He is now with Institute of Micro- and Nanoelectronic Systems, Karlsruhe Institute of Technology.

N. Karcher, O. Sander, M. Weber and T. Wolber are with Institute for Data Processing and Electronics, Karlsruhe Institute of Technology.

C. Enss is with Kirchhoff Institute for Physics, Heidelberg University, and Institute for Data Processing and Electronics, Karlsruhe Institute of Technology.

S. Kempf was with Kirchhoff Institute for Physics, Heidelberg University. He is now with Institute of Micro- and Nanoelectronic Systems and Institute for Data Processing and Electronics, both Karlsruhe Institute of Technology. (sebastian.kempf@kit.edu)

Color versions of one or more figures in this paper are available online at http://ieeexplore.ieee.org.
}}

\markboth{Presentation Number: 1EOr2A-02}%
{Demonstration of simultaneous MMC readout using a tailored microwave SQUID multiplexer based readout system}

\maketitle

\begin{abstract}
Magnetic microcalorimeters (MMCs) are cryogenic, energy-dispersive single-particle detectors providing excellent energy resolution, intrinsically fast signal rise time, quantum efficiency close to 100\%, large dynamic range as well as almost ideal linear response. One of the remaining challenges to be overcome to ultimately allow for the utilization of large-scale MMC based detector arrays with thousands to millions of individual pixels is the realization of a SQUID based multiplexing technique particularly tailored for MMC readout. Within this context, we report on the first demonstration of a frequency-division multiplexed readout of an MMC based detector array using both, a custom microwave SQUID multiplexer as well as a dedicated software-defined radio (SDR) readout electronics. We successfully performed a simultaneous readout up to eight multiplexer channels, each monitoring two detector pixels. We show that the signal shape is not changed as compared to a dc-SQUID readout and that similar values for the energy resolution can be obtained. Nevertheless, we observed an influence of the internal quality factor of the microwave resonators used for frequency encoding on the energy resolution that affects the resolution of the co-added sum spectrum. 
\end{abstract}


\section{Introduction}
Progress in the understanding of nature is often accompanied by advances in physical instrumentation. For this reason, intensive R\&D efforts are routinely taken to push detector technology beyond the present state of the art and to challenge physical and technological limitations. Within this context, magnetic microcalorimeters (MMCs) \cite{Kem18, Fle09} have proven to outperform conventional detectors by orders of magnitude and allow performing experiments that have been considered impossible so far. 

MMCs are energy-dispersive, cryogenic single-particle detectors. They consist of an absorber for the particles to be detected that is weakly coupled to a heat bath kept at constant temperature $T_\mathrm{bath}$. The absorber is strongly coupled to a paramagnetic temperature sensor situated in a weak external magnetic field to induce a temperature-dependent sensor magnetization $M(T)$. The latter is altered by the temperature rise $\delta T = \delta E / C_\mathrm{det}$ of the detector upon the deposition of the energy $\delta E$ within the absorber. Here, $C_\mathrm{det}$ denotes the total heat capacity of the detector. For readout, the magnetization change $\delta M \propto (\partial M/\partial T) \delta T$ is sensed as a change of magnetic flux $\delta \Phi \propto \delta M$ threading a superconducting pickup coil that is inductively coupled to the temperature sensor and connected to the input coil of a current-sensing SQUID to form a superconducting flux transformer. In this configuration, the change of screening current $\delta I_\mathrm{scr} \propto \delta \Phi$ within the flux transformer resulting from the magnetization change of the temperature sensor is transduced into a change of the output signal of the SQUID.

Single-channel MMCs as well as small-scale detector arrays are typically read out by individual current-sensing dc-SQUIDs in a two-stage SQUID readout configuration with high-speed FLL electronics \cite{Kem18}. Though this readout approach is well-established and enabled impressive demonstrations of the outstanding performance of MMCs (see, for example, \cite{Sik20, Bat15, Ran17}), it can hardly be scaled up for large-scale detector arrays as the number of readout electronics, cost, parasitic power dissipation and system complexity is linearly increasing with the number of detector channels. Consequently, cryogenic SQUID-based multiplexing techniques are presently being developed, among which microwave SQUID multiplexing ($\mu$MUXing) \cite{Irw04, Mat08, Kem17} appears to be most promising.

Each channel of a microwave SQUID multiplexer consists of a non-hysteretic, current-sensing rf-SQUID that is inductively coupled to a superconducting microwave resonator via a superconducting load inductor. Due to the magnetic flux dependence of the SQUID inductance as well as the mutual interaction between the SQUID and the associated resonator, the detector signal is transduced into a resonance frequency shift of the resonator. For actual multiplexing, $N$ individual resonance circuits, each having a unique resonance frequency, are capacitively coupled to a common transmission line (feedline). In this configuration, a continuous-wave microwave frequency comb is used to probe the resonators altering the amplitude and phase of the transmitted carriers according to the actual resonance frequency and hence the detector signal. Here, creation of the frequency comb and online analysis of the amplitude and phase changes is achieved by using a software defined radio (SDR) system operated at room temperature \cite{Ker18, Kar20}. For output signal linearization, flux ramp modulation is used \cite{Leh07, Mat12}. For this, a periodic, sawtooth-shaped current signal is injected into a common modulation coil connected to all readout SQUIDs to induce a linearly rising flux ramp with an amplitude of several flux quanta inside the SQUID loop. The latter continuously varies the SQUID inductance (and hence the resonance frequency of the associated resonator) according to the SQUID characteristic. The flux ramp repetition rate sets the effective sampling rate and, hence, the signal bandwidth. It is chosen such that the input signal appears to be quasi-static within a flux ramp cycle. In this mode, an input signal leads to a constant magnetic flux offset causing a phase-shift of the SQUID characteristic that depends linearly on the actual amplitude of the input signal. 

To actually prove the suitability of this multiplexing technique for MMC readout, we have developed a tailored microwave SQUID multiplexer based readout system and report on the first demonstration of a multiplexed readout of an MMC based detector array using both, a custom microwave SQUID multiplexer as well as a dedicated software-defined radio (SDR) readout electronics.

\section{Multiplexer description}
Our readout system consists of a custom cryogenic microwave SQUID multiplexer, particularly developed for MMC readout to guarantee impedance matching and minimization of stray inductances, as well as a custom software-defined radio (SDR) based readout electronics matched to our cryogenic multiplexer \cite{Kar20, San19}. As the hardware and firmware configuration of the latter is a rather complex subject on its own and actually not the focus of this paper, we omit a description here and refer to the cited reference for a detailed and appropriate discussion.

\begin{figure}[!t]
\centering
\includegraphics[width=1.0\columnwidth]{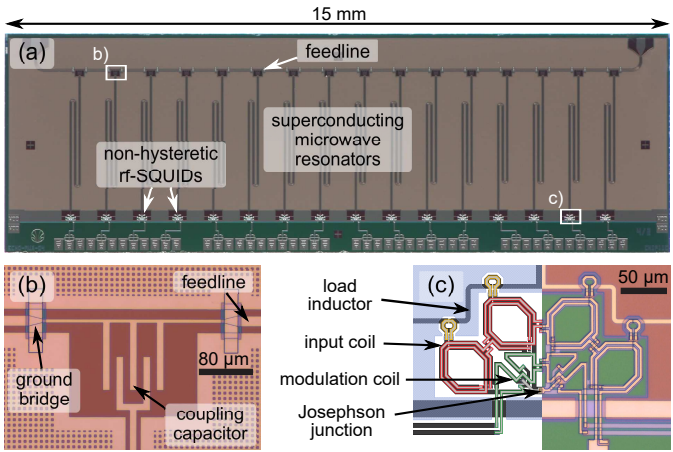}
\caption{(a) Photograph of the microwave SQUID multiplexer prototype fabricated for this work. (b) Microscope photograph of the region around the coupling capacitor through which each superconducting microwave  resonator is capacitively coupled to the feedline. (c) Layout (left) and microscope photograph (right) of one of the non-hysteretic rf-SQUIDs that are inductively coupled to the resonators.}
\label{fig_1}
\end{figure}

The microwave SQUID multiplexer developed for this work is shown in Fig.~\ref{fig_1}(a) and is based on sixteen coplanar waveguide (CPW) superconducting quarterwave transmission line resonators that are capacitively coupled to a CPW feedline. Assuming negligible internal losses within the resonators, i.e. $Q_\mathrm{i} \rightarrow \infty$ with $Q_\mathrm{i}$ being the internal quality factor, we chose the coupling capacitors (see Fig.~\ref{fig_1}(b)) such that the bandwidth of each readout channel is about $1\,\mathrm{MHz}$ to match to the signal rise time of the ECHo detector array \cite{Man22} that we used for this work. We varied the length of the resonators such that the resonance frequency of bare resonators, i.e. without load inductor and coupling capacitor, are within the frequency band $4.5\,\mathrm{GHz}$ and $4.8\,\mathrm{GHz}$. The frequency spacing between neighboring resonators is $20\,\mathrm{MHz}$ and will be later reduced to about $10\,\mathrm{MHz}$ for close packing density while simultaneously keeping the crosstalk between resonators at a negligible level \cite{Mat19}.

Each resonator is terminated with a superconducting load inductor that is connected to ground at one end and to the CPW resonator center line at the other end. Load inductor and rf-SQUID are inductively coupled (see Fig.~\ref{fig_1}(c)). The mutual inductance $M_\mathrm{T}$ between SQUID and resonator is $3.8\,\mathrm{pH}$ to yield a peak-to-peak resonance frequency shift of $0.7 \ldots 0.8\,\mathrm{MHz}$ depending on the actual resonance frequency. The non-hysteretic rf-SQUID (see Fig.~\ref{fig_1}(c)) is based on a parallel connection of four SQUID loops, each having an inductance of about $180\,\mathrm{pH}$. The critical current of the Nb/Al-AlO$_{\mathrm{x}}$/Nb-Josephson tunnel junction within each SQUID is $4.3\,\mu\mathrm{A}$, resulting in a SQUID hysteresis parameter $\beta_\mathrm{L} = 2\pi L_\mathrm{s} I_\mathrm{c}/\Phi = 0.6.$ For coupling to the detector, the SQUID is equipped with an inductively coupled superconducting input coil having an inductance of about $1.6\,\mathrm{nH}$. It is hence impedance-matched to ECHo detector array \cite{Man22}. To reduce microwave leakage into the detector, we connected a resistor with a resistance of $3\,\Omega$ in parallel to the input coil to form a first-order $LR$-low-pass filter within the input circuit. It is worth mentioning that all inductances and mutual inductances get reduced when connecting the MMC pickup coil due to the screening of the resulting superconducting flux transformer.

\section{Experimental setup}

We mounted the in-house fabricated multiplexer prototype together with a 64 pixels ECHo detector array onto a sample holder made of copper and cooled down the setup to about $T_\mathrm{bath} = 10\,\mathrm{mK}$ using a commercial $^3$He/$^4$He dilution refrigerator with pulse tube cooler based pre-cooling circuit. We connected our custom SDR based readout electronics to the multiplexer using flexible coaxes outside and semi-rigid coaxes within the cryostat. We attenuated the multiplexer input line by $33\,\mathrm{dB}$ using cryogenic attenuators and directional couplers thermally anchored at different temperature stages of the cryostat to set the required power level of the readout tones while minimizing thermal noise at the same time. For boosting the tiny multiplexer output signal, i.e. the modulated microwave frequency comb, we connected a high-gain, ultra-low noise HEMT amplifier mounted at the $4\,\mathrm{K}$ temperature stage of the cryostat via a superconducting coaxial cable to the multiplexer output port.

\begin{figure}[!t]
\centering
\includegraphics[width=0.8\columnwidth]{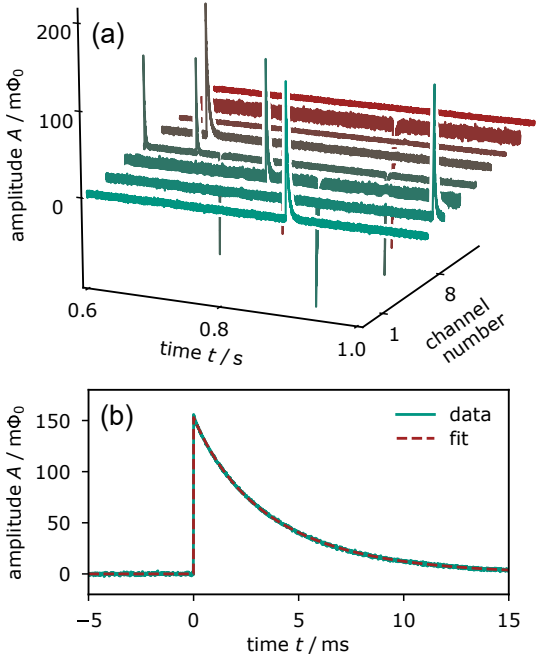}
\caption{(a) Demodulated time streams of the individual multiplexer readout channels for a measurement in which eight multiplexer channels are simultaneously operated. Each channel is connected to a two-pixel MMC whereas the different pixels of a single detector can be distinguished by the signal polarity. (b) Single detector signal occurring in one of the multiplexer readout channels. In addition, a fit to the data is shown that is based on the thermodynamic model of an MMC yielding the time constants of the detector. The latter correspond to our expectations that are based on multiplexer and detector parameters.}
\label{fig_2}
\end{figure}

After performing some sanity checks and measuring the basic characteristics of the microwave SQUID multiplexer without putting the detector array in operation, we biased the array by injecting a persistent current of $I_\mathrm{F} = 30\,\mathrm{mA}$ into the pickup coils to create a bias magnetic field and irradiated the different pixels of the detector array with X-ray photons emitted by an $^{55}$Fe calibration source. It is worth mentioning that we chose the value of the persistent current according to detector optimization calculations. For flux ramp modulation, we injected a sawtooth shaped flux ramp signal with an amplitude of $3.4\,\Phi_0$ and a repetition rate of $125\,\mathrm{kHz}$ into the common modulation line. We trimmed the digitized multiplexer output time stream per flux ramp cycle to cut out transient regions, ultimately yielding a flux ramp signal with an amplitude of $3\,\Phi_0$ per cycle. For demodulation of the flux ramp modulated signal, we used the method described in \cite{Mat12}. We measured individual detector signals and investigated whether the signal shape changed as compared to the readout using individual dc-SQUIDs and then measured several spectra of the calibration source varying, for example, the number of simultaneously operated readout channels. 

\section{Results and discussion}

For checking whether the detectors are affected by the microwave readout tones and whether a significant power dissipation within the multiplexer potentially prevents the detector from cooling down to base temperature, we varied the microwave power of the readout tones and measured for each value the temperature dependence $M(T)$ of the sensor magnetization by varying the heat bath temperature. All measured curves are lying on top of each other, thus the magnetization curve is not affected by the power of the microwave carriers. Moreover, comparing the shape of the magnetization curve with predictions based on Monte Carlo simulations \cite{Fle05}, we see that the chip temperature is very close to the heat bath temperature. For this reason, we tried to use the optimal readout power for each readout channel for all subsequent measurements. However, we figured out that the present combination of the analog front-end of the SDR readout electronics and the cryogenic microwave setup does not allow to apply optimum readout power when increasing the channel count as the finite DAC output range is limited and must be shared among all readout tones. For this reason, we plan to implement another microwave amplifier behind the DAC in the next version of the readout electronics.

\begin{figure}[!t]
\centering
\includegraphics[width=1.0\columnwidth]{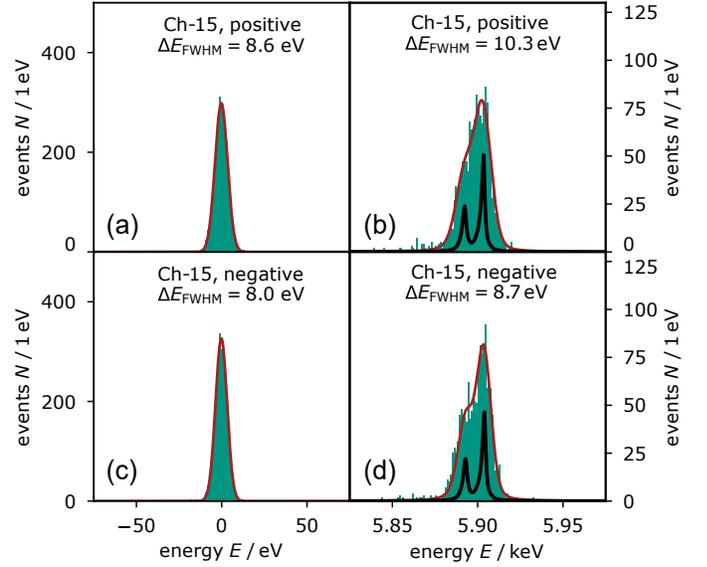}
\caption{Histograms of the measured energy spectra of baseline (a,c) and detector (b,d) signals upon the deposition of K$_\alpha$-photons emitted from the utilized $^{55}$Fe calibration source for both detector pixels of readout channel 15. The red solid lines are fits to the measured histograms assuming a Gaussian detector response with an intrinsic detector resolution $\Delta E_\mathrm{FWHM}$. In addition, the black solid line denotes the natural line shape of the K$_\alpha$-line in (b,d).}
\label{fig_3}
\end{figure}

Fig.~\ref{fig_2} shows the measured time streams of the individual detector pixels when simultaneously reading out eight multiplexer channels (and hence sixteen detector pixels) within a time window of $400\,\mathrm{ms}$ length as well as a zoom into one of the signals. As expected, the overall signal shape and signal height is unaltered as compared to single channel dc-SQUID readout. The slight deviation of the signal height of the positive and negative pulses are due to inductance asymmetry and fabrication tolerances (sensor thickness and area) within each detector. This plot nicely shows that our multiplexing technique works as intended and allows to simultaneously read out independent detector channels.

Fig.~\ref{fig_3} shows, as an example, the energy spectrum of multiplexer channel 15 when simultaneously operating four multiplexer readout channels. The baseline energy resolution is about $8\,\mathrm{eV}$ and the resolution at $6\,\mathrm{keV}$ is about $8-10\,\mathrm{eV}$. The slight difference between positive and negative pulses is due to the difference in signal height (see above). The measured resolution is in accordance with the measured detector and multiplexer characteristics and can be reduced in future by optimizing the multiplexer and detector parameters.

\begin{figure}[!t]
\centering
\includegraphics[width=1.0\columnwidth]{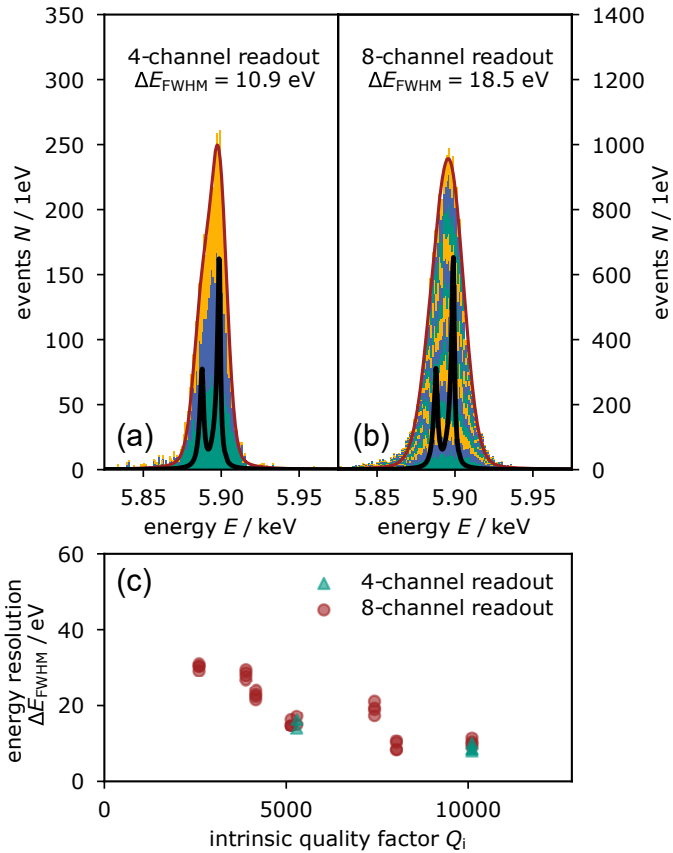}
\caption{Energy spectra after co-adding all acquired detector signals when simultaneously operating (a) four and (b) eight multiplexer readout channels. The solid lines are fits to the histogram assuming a Gaussian detector response with an intrinsic detector resolution $\Delta E_\mathrm{FWHM}$ as given in the plots. In addition, the natural lineshape of the K$_\alpha$-line is shown. c) Dependence of the measured energy resolution of the individual signal channels on the internal quality factor of the microwave resonators when simultaneously operating four and eight readout channels, respectively.}
\label{fig_4}
\end{figure}

Fig.~\ref{fig_4} shows the co-added spectra when simultaneously operating (a) four and (b) eight multiplexer channels. Obviously, two things are striking about the histograms. First, the determined energy resolution of the co-added spectra is worse as compared to Fig.~\ref{fig_3}. Second, the resolution degrades when increasing the number of channels that are simultaneously operated.

The degradation of the energy resolution of different channels within a single measurement ultimately causing the deterioration of the co-added spectrum results from a systematic dependence of the energy resolution on the internal quality factor of the resonators. For illustration, Fig.~\ref{fig_4}(c) shows the dependence of the energy resolution determined for each channel within the 4/8-channel measurement on the internal quality factor. Obviously, the energy resolution degrades the lower the internal quality factor is. A similar observation was made in \cite{Hir13, Koh14}, where a dependence of the readout noise of two different SQUID multiplexers on the intrinsic quality factor of the used resonators is found, too.

The reason for this behavior is presently not fully resolved. We observe a strong degradation of the internal quality factor when connecting the detector array to the external bias lines. This points towards an influence of the external detector bias circuitry. More precisely, it presently seems that resonators with a small distance to the external bias lines have a lower internal quality factor than resonators with a large distance. However, this explanation is not secured and we are presently carrying out systematic investigations of this issue. But as we regularly measure internal quality factors $Q_\mathrm{i}>5\times 10^4$ at $\mathrm{mK}$ temperatures, we can already state that the internal quality factors are fine when no detector array is connected to the multiplexer, i.e. it is more an issue related to the detector arrangement than the multiplexing technique.

The difference of energy resolution between individual channels due to the variation of the intrinsic quality factor is the main reason why the energy resolution of the sum spectrum of the 8-channel measurement is worse than of the 4-channel measurement as more channels with worse resolution are added up. Moreover, the energy resolution for all channels is in general slightly reduced due to the decrease of the readout power per carrier signal  when increasing the number of readout channels . This results from the present limitation of our SDR based readout electronics (see above) and will hence be somewhat  easily resolved by using the next version.

For estimating the crosstalk between neighboring resonators, we used a similar technique as described in \cite{Mat19}. We acquired the data stream of a victim resonator channel and used  the detector signal of a perpetrator readout channel as trigger. By relating the signal amplitude for both readout channels (ideally the amplitude of the victim channel should be zero), we estimated the upper crosstalk limit for different resonator pairs and found to be well below $1\,\%$. This is as expected as the frequency spacing between neighboring channels is rather large (see above) and the spatial distribution of the resonators is adapted for low crosstalk levels. Systematic measurements using another multiplexer prototype moreover show that even for a $10\,\mathrm{MHz}$ frequency spacing the crosstalk level is low enough to guarantee for MMC readout without energy resolution deterioration. 

\section{Conclusions}
In conclusion, we presented the first demonstration of a multiplexed readout of an MMC based detector array using both, a custom microwave SQUID multiplexer as well as a dedicated software-defined radio (SDR) readout electronics. More precisely, we showed that this multiplexing technique is very well suited for MMCs as the signal shape is not degraded as compared to conventional single channel dc-SQUID readout and similar values of the energy resolution can be obtained. Some open challenges remain, e.g. improving the internal quality factor of the microwave resonators when a detector is connected to the related microwave SQUID. Nevertheless, this demonstration paves the way for the realization of large-scale and ultralarge-scale MMC based detector arrays as foreseen, for example, for the ECHo experiment.

\balance

\bibliographystyle{IEEEtran}
\bibliography{literature}

\end{document}